\documentclass[12pt,british]{iopart}
\usepackage[T1]{fontenc}
\usepackage[latin9]{inputenc}
\usepackage{graphicx}
\usepackage{amssymb}

\makeatletter

\providecommand{\boldsymbol}[1]{\mbox{\boldmath $#1$}}

\usepackage{iopams}
\usepackage{setstack}

\usepackage{iopams}

\def\vec#1{\mbox{\boldmath $#1$}}
\def\Rm{\mathit{Rm}}
\def\RE{\mathit{Re}}
\def\d{\rmd}
\def\i{\rmi}

\usepackage{babel}
\makeatother

\begin{document}

\title{Contactless Electromagnetic Phase-Shift Flowmeter for Liquid Metals}

\author{J\={a}nis Priede$^{1}$, Dominique Buchenau$^{2}$ and Gunter Gerbeth$^{2}$}

\address{$^{1}$Applied Mathematics Research Centre, Coventry University,
United Kingdom}

\address{$^{2}$Forschungszentrum Dresden-Rossendorf, MHD Department, Germany}

\ead{j.priede@coventry.ac.uk}

\begin{abstract}
We present a concept and test results of an eddy-current flowmeter
for liquid metals. The flow rate is determined by applying a weak
ac magnetic field to a liquid metal flow and measuring the flow-induced
phase disturbance in the external electromagnetic field. The phase
disturbance is found to be more robust than that of the amplitude
used in conventional eddy-current flowmeters. The basic characteristics
of this type of flowmeter are analysed using simple theoretical models,
where the flow is approximated by a solid body motion. Design of such
a flowmeter is presented and its test results reported. 
\end{abstract}

\noindent{\it Keywords\/}: {\noindent Electromagnetic flowmeter, liquid metal, ac magnetic field }

\pacs{41.20.Gz, 47.60.Dx, 47.65.-d, 47.80.Cb}

\submitto{\MST}

\maketitle

\section{\label{Intro} Introduction}

Accurate and reliable flow rate measurements are required in many
technological processes using liquid metals. Commercially available
electromagnetic flowmeters typically use electrodes in direct contact
to the liquid to measure the voltage induced by the flow in dc magnetic
field \cite{JAS,Bevir,Engl_2,Shimizu}. The use of electrodes is problematic
in aggressive media like molten metals, for which a contactless treatment
is preferable \cite{Duncombe}. A well-known example of such a contactless
flowmeter is the magnetic flywheel, which is described in the textbook
of Shercliff \cite{JAS} and employed by Bucenieks \cite{Buc1,Buc2}
for the flow rate measurements. Such kind of flowmeters have recently
been reembodied under the name of Lorentz force velocimetry \cite{Thess1,Thess2}.
As the name suggests, the Lorentz force sensors measure the electromagnetic
force exerted by the flow on a closely placed permanent magnet. This
force is proportional to the product of the electrical conductivity
of the fluid and the square of the applied magnetic field strength.
Another type of flowmeter using a single rotating magnet has recently
been reported in \cite{Priede1,PBG10}. This sensor is based on the
equilibrium of the electromagnetic torques caused by the flow on the
magnet and by the magnet on the flow. The equilibrium rotation rate
is, in a reasonable range of parameters, independent of both the strength
of the permanent magnet and the conductivity of the liquid, which
makes the measurements insensitive to temperature variations in the
liquid.

Alternatively, the flow of liquid metal can be determined in a contactless
way by eddy-current flowmeters, which measure the flow-induced perturbation
of an externally applied magnetic field \cite{LL48,Cowley65}. This
principle underlies also the so-called flow tomography approach using
either dc \cite{Stefa} or ac magnetic fields \cite{Gund}. The main
problem of this method is to measure a weak induced magnetic field
with the relative amplitude of the order of magnitude of the magnetic
Reynolds number $\Rm\sim10^{-4}-10^{-1}$ on the background of the
applied magnetic field. There are a number of measurement schemes
known which rely on the geometrical compensation of the applied field
by a proper arrangement of sending and receiving coils so that only
the signal induced by the flow is measured \cite{Feng,Dement}. Such
flowmeters employ the flow-induced asymmetry of the magnetic field.
Unfortunately, there are a number of side effects such as, for example,
the thermal expansion, which can also cause some asymmetry between
the receiving coils.

As the flow can disturb not only the amplitude of an ac magnetic field
but also its phase distribution, the latter can also be used for the
flow rate measurements \cite{Sendai,Pat}. In this paper, we analyse
the basic characteristics of such a phase-shift flowmeter, present
its technical implementation and report the test results. 

The paper is organised as follows. The basic physical effects are
considered in Section \ref{model} using a simple model where the
liquid flow is approximated by a solid body motion. In Section \ref{Realization}
we describe the realization of the phase-shift sensor and present
flow rate measurements at two different liquid metal loops. The paper
is concluded by a summary in Section \ref{Sum}.

\section{\label{model}Mathematical model}

\subsection{\label{eqs}Basic equations}

Consider a medium of electrical conductivity $\sigma$ moving with
the velocity $\vec{v}=\vec{e}_{x}V$ in an ac magnetic field with
the induction $\vec{B}$ alternating harmonically with the angular
frequency $\omega.$ The induced electric field follows from the Maxwell-Faraday
equation as $\vec{E}=-\vec{\nabla}\Phi-\partial_{t}\vec{A},$ where
$\Phi$ is the electric potential, $\vec{A}$ is the vector potential
and $\vec{B}=\vec{\nabla}\times\vec{A}$. The density of the electric
current induced in the moving medium is given by Ohm's law \[
\vec{j}=\sigma(\vec{E}+\vec{v}\times\vec{B})=\sigma(-\vec{\nabla}\Phi-\partial_{t}\vec{A}+\vec{v}\times\vec{\nabla}\times\vec{A}).\]
 Assuming the ac frequency to be sufficiently low to neglect the displacement
current, Ampere's law $\vec{j}=\frac{1}{\mu_{0}}\vec{\nabla}\times\vec{B}$
leads to the following advection-diffusion equation for the vector
potential

\begin{equation}
\partial_{t}\vec{A}+(\vec{v}\cdot\vec{\nabla})\vec{A}=\frac{1}{\mu_{0}\sigma}\nabla^{2}\vec{A},\label{eq:A-gen}\end{equation}
 where the gauge invariance of $\vec{A}$ has been used to specify
the scalar potential as \[
\Phi=\vec{v}\cdot\vec{A}-\frac{1}{\mu_{0}\sigma}\vec{\nabla}\cdot\vec{A}.\]
 In the following, we consider an applied magnetic field varying in
time harmonically as $\vec{A}_{0}(\vec{r},t)=\vec{A}_{0}(\vec{r})\cos(\omega t),$
which allows us to search for a solution in the complex form $\vec{A}(\vec{r},t)=\Re\left[\vec{A}(\vec{r})e^{\i\omega t}\right]$.
Then equation (\ref{eq:A-gen}) for the amplitude distribution of
the vector potential takes the form

\begin{equation}
\i\omega\vec{A}+(\vec{v}\cdot\vec{\nabla})\vec{A}=\frac{1}{\mu_{0}\sigma}\nabla^{2}\vec{A}.\label{eq:A-amp}\end{equation}
Further we focus on a simple 2D externally applied magnetic field,
which is invariant along the unit vector $\vec{\epsilon}.$ Such a
magnetic field can be specified by a single component of the vector
potential $\vec{A}=\vec{\epsilon}A$ as $\vec{B}=\vec{\nabla}\times\vec{\epsilon}A=-\vec{\epsilon}\times\vec{\nabla}A,$
where $\vec{B}$ has only two components in the plane perpendicular
to $\vec{\epsilon}.$ The continuity of $\vec{B}$ at the interface
$S$ between conducting and insulating media imply the following boundary
conditions: \begin{equation}
\left[A\right]_{S}=\left[\partial_{n}A\right]_{S}=0,\label{eq:bcnd}\end{equation}
 where $\left[f\right]_{S}$ denotes the jump of quantity $f$ across
the boundary $S$; $\partial_{n}\equiv(\vec{n}\cdot\vec{\nabla})$
is the derivative normal to the boundary.

\subsection{Solution for a single harmonic of the magnetic field}

\label{layer} 

\begin{figure*}
\begin{centering}
\includegraphics[width=0.5\columnwidth]{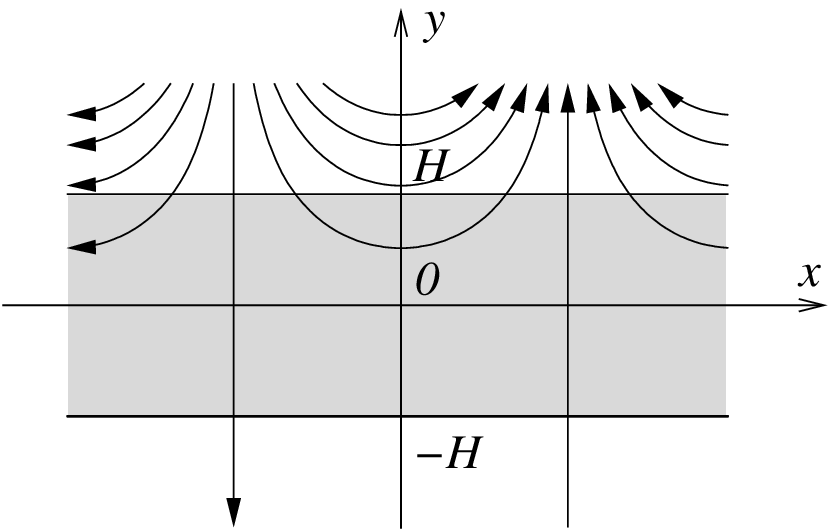}{\put(-135,5){(\textit{a})}}\includegraphics[width=0.5\columnwidth]{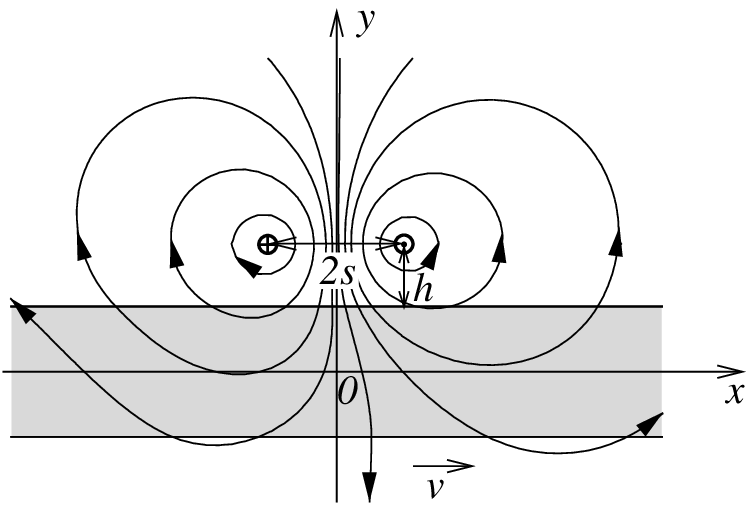}{\put(-140,5){(\textit{b})}}
\par\end{centering}

\caption{\label{cap:sketch}Model of a conducting layer of thickness $2H$
in an external magnetic field represented by a standing harmonic wave
(\emph{a}) and generated by a couple of straight wires (\emph{b}).}

\end{figure*}

We start with a simple model shown in figure \ref{cap:sketch}(a),
where the conducting medium is a layer of thickness $2H,$ and the
applied magnetic field is a harmonic standing wave with the vector
potential amplitude given by \[
\vec{A}_{0}(\vec{r},t)=\vec{e}_{z}A_{0}(\vec{r},t)=\vec{e}_{z}\hat{A}_{0}(y)\cos(kx)\cos(\omega t),\]
 where $k$ is the wavenumber in the $x$-direction. Henceforth, we
choose the half-thickness $H$ as the length scale and introduce a
dimensionless ac frequency and the magnetic Reynolds number, \begin{eqnarray}
\bar{\omega} & = & \mu_{0}\sigma\omega H^{2},\label{eq:omegb}\\
\Rm & = & \mu_{0}\sigma VH,\label{eq:Rm}\end{eqnarray}
where the latter represents a dimensionless velocity. It is important
to note that this key parameter depends on the product of the physical
velocity and electrical conductivity. Although for a typical liquid
metal flow $\Rm\ll1,$ the following analysis will not be restricted
to this case unless stated otherwise. For a free space, where $\sigma=0,$
equation (\ref{eq:A-amp}) takes the form \begin{equation}
\frac{\d^{2}\hat{A}_{0}}{\d y^{2}}-k^{2}\hat{A}_{0}=0,\label{eq:A0}\end{equation}
 and has the solution \begin{equation}
\hat{A}_{0}(y;k)=C_{0}e^{|k|(y-1)},\label{sol:A0}\end{equation}
 where the constant \begin{equation}
C_{0}=\hat{A}_{0}(1;k)\label{eq:C0}\end{equation}
defines the amplitude of the Fourier mode with the wavenumber $k$
of the external magnetic field at the upper boundary of the layer.
It is important to note that the external magnetic field, which is
assumed in the form of a standing wave, can be represented as a superposition
of two oppositely travelling waves \[
A_{0}(\vec{r},t)=\frac{1}{2}\left[A_{0}^{+}(\vec{r},t)+A_{0}^{-}(\vec{r},t)\right],\]
 where $A_{0}^{\pm}(\vec{r},t)=\hat{A}_{0}(y)\cos(\omega t\pm kx).$
This implies that the solution can be sought in a similar form as
\[
A(\vec{r},t)=\frac{1}{2}\left[A^{+}(\vec{r},t)+A^{-}(\vec{r},t)\right],\]
where $A^{\pm}(\vec{r},t)=\Re\left[\hat{A}(y;\pm k)e^{\i(\omega t\pm kx)}\right]$
are oppositely travelling fields. The solution governed by equation
(\ref{eq:A0}) in the free space above the layer $(y\ge1),$ can be
written as \begin{equation}
\hat{A}(y;k)=\hat{A}_{0}(y;k)+\hat{A}_{1}(y;k),\label{sol:A-a}\end{equation}
 where the first term represents the external field (\ref{sol:A0})
and $\hat{A}_{1}(y;k)=C_{1}e^{-|k|(y-1)}$ is the induced field. In
the free space below the layer, $y\le-1$, the solution satisfying
(\ref{eq:A0}) is \begin{equation}
\hat{A}(y;k)=C_{3}e^{|k|(y+1)}.\label{sol:A-b}\end{equation}
 In the conducting layer, $-1\le y\le1$, equation (\ref{eq:A-amp})
for a travelling field takes the form \begin{equation}
\frac{\d^{2}\hat{A}}{\d y^{2}}-\kappa^{2}\hat{A}=0,\label{eq:A-i}\end{equation}
 where $\kappa(k)=\sqrt{k^{2}+\i(\bar{\omega}+k\Rm})$, and has the
solution \begin{equation}
\hat{A}(y;k)=C_{2}\sinh(\kappa y)+D_{2}\cosh(\kappa y).\label{sol:A-i}\end{equation}
The unknown constants, which are regarded as functions of the wavenumber
$k,$ are found from the boundary conditions (\ref{eq:bcnd}) as follows\begin{eqnarray}
C_{2} & = & C_{0}|k|/\left(|k|\sinh(\kappa)+\kappa\cosh(\kappa)\right)\label{eq:C2}\\
D_{2} & = & C_{0}|k|/\left(|k|\cosh(\kappa)+\kappa\sinh(\kappa)\right),\\
C_{1} & = & D_{2}\cosh(\kappa)+C_{2}\sinh(\kappa)-C_{0},\\
C_{3} & = & D_{2}\cosh(\kappa)-C_{2}\sinh(\kappa).\label{eq:C3}\end{eqnarray}

\subsection{Solution for an external magnetic field generated by a couple of
straight wires}

The solution above can easily be extended to an external magnetic
field generated by a finite-size coil. The simplest model of such
a coil consists of two parallel straight wires fed with an ac current
of amplitude $I_{0}$ flowing in the opposite directions along the
$z$-axis at distance $2s$ in the $x$-direction and placed at height
$h$ above the upper surface of the layer, as shown in figure \ref{cap:sketch}(b).
The free-space distribution of the vector potential amplitude having
only the $z$-component, which is further scaled by $\mu_{0}I_{0},$
is governed by \begin{equation}
\vec{\nabla}^{2}A_{0}=-\delta(\vec{r}-h\vec{e}_{y}-s\vec{e}_{x})+\delta(\vec{r}-h\vec{e}_{y}+s\vec{e}_{x}),\label{eq:A02}\end{equation}
where $\delta(\vec{r})$ is the Dirac delta function and $\vec{r}$
is the radius vector. The problem is solved by the Fourier transform
$\hat{A}(y;k)=\int_{-\infty}^{\infty}A(x,y)e^{\i kx}\,\d x,$ which
converts (\ref{eq:A02}) into

\begin{equation}
\frac{\d^{2}\hat{A}_{0}}{\d y^{2}}-k^{2}\hat{A}_{0}=-f(k)\delta(y-h),\label{eq:A0-k}\end{equation}
 where $f(k)=\int_{-\infty}^{\infty}[\delta(x-s)-\delta(x+s)]e^{\i kx}\,\d x=2\i\sin(ks).$
The solution of (\ref{eq:A0-k}) decaying at $y\rightarrow\pm\infty$
can be written as \begin{equation}
\hat{A}_{0}(y;k)=c(k)e^{-|k(y-h)|},\label{sol:A0-k}\end{equation}
 where the unknown coefficient \begin{equation}
c(k)=\frac{f(k)}{2|k|}=\frac{\i\sin(ks)}{|k|}.\label{eq:ck}\end{equation}
follows from the boundary condition $\left.\left[\d\hat{A}/\d y\right]\right|_{y=h}=-f(k),$
which is obtained by integrating equation (\ref{eq:A0-k}) over the
singularity at $z=h$. The solution for a separate Fourier mode in
the regions above, below and inside the layer is given, respectively,
by expressions (\ref{sol:A-a}, \ref{sol:A-b}) and (\ref{sol:A-i})
with the coefficients (\ref{eq:C2})--(\ref{eq:C3}) containing the
constant $C_{0},$ which is defined by substituting (\ref{sol:A0-k})
into (\ref{eq:C0}). Then the spatial distribution of the complex
vector potential amplitude is given by the sum of Fourier modes, which
is defined by the inverse Fourier transform $A(x,y)=\frac{1}{2\pi}\int_{-\infty}^{\infty}\hat{A}(y;k)e^{-\i kx}\,\d k$
and can efficiently be calculated using the Fast Fourier Transform.

\subsection{\label{numres}Numerical results}

\noindent For physical interpretation of the following results, note
that the magnetic flux through a surface is given by the circulation
of the vector potential along the contour encircling that surface.
For the simple 2D case under consideration, when the vector potential
has only one component, the difference of the vector potential between
two points defines the linear flux density between two lines parallel
to the vector potential at those two points. The same holds also for
the time derivative of the corresponding quantities. Thus, the difference
of the vector potential amplitudes between two points is proportional
to the e.m.f. amplitude which could be measured by an idealised coil
consisting of two straight parallel wires placed along the $z$-axis
at those points. Correspondingly, the single-point vector potential
considered below gives the e.m.f. measured by a `wide' coil with the
second wire placed sufficiently far away in the region of a negligible
magnetic field.

\subsubsection{Single harmonic of the magnetic field}

\begin{figure*}
\begin{centering}
\includegraphics[bb=125bp 90bp 355bp 250bp,clip,width=0.49\columnwidth]{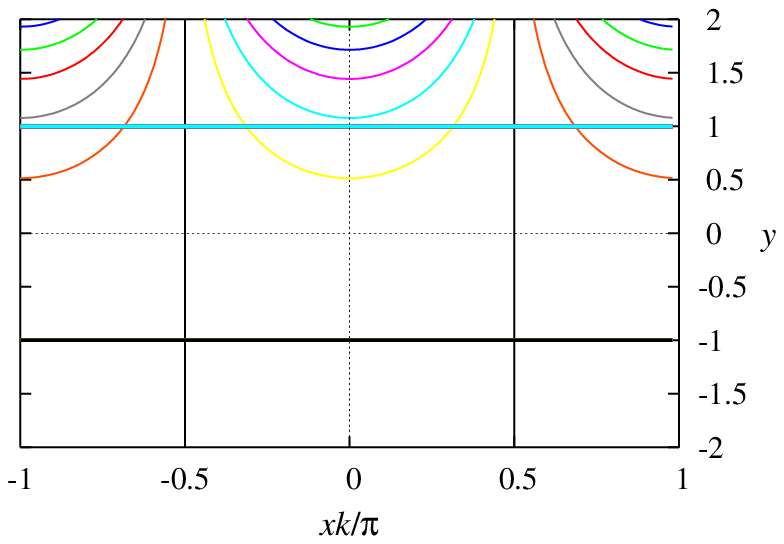}\put(-45,30){(\textit{a})}\includegraphics[bb=125bp 90bp 355bp 250bp,clip,width=0.49\columnwidth]{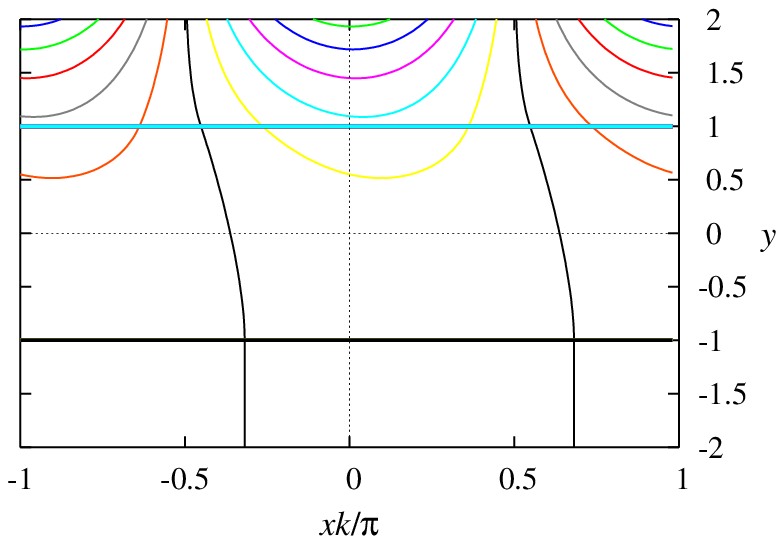}\put(-45,30){(\textit{b})}
\par\end{centering}

\begin{centering}
\includegraphics[bb=125bp 90bp 355bp 250bp,clip,width=0.49\columnwidth]{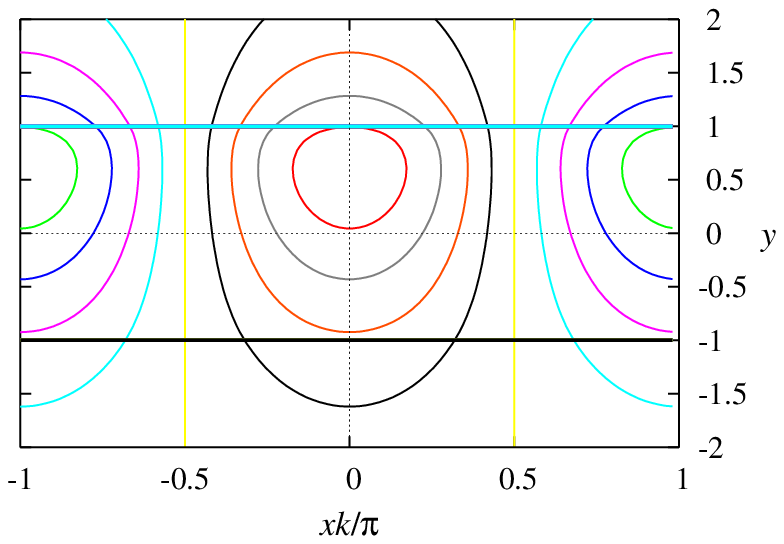}\put(-45,30){(\textit{c})}\includegraphics[bb=125bp 90bp 355bp 250bp,clip,width=0.49\columnwidth]{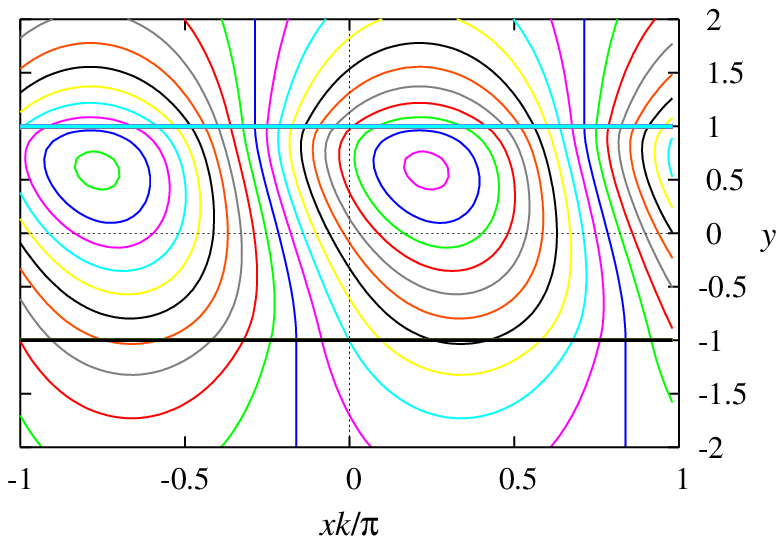}\put(-45,30){(\textit{d})}
\par\end{centering}

\begin{centering}
\includegraphics[bb=125bp 90bp 355bp 250bp,clip,width=0.49\columnwidth]{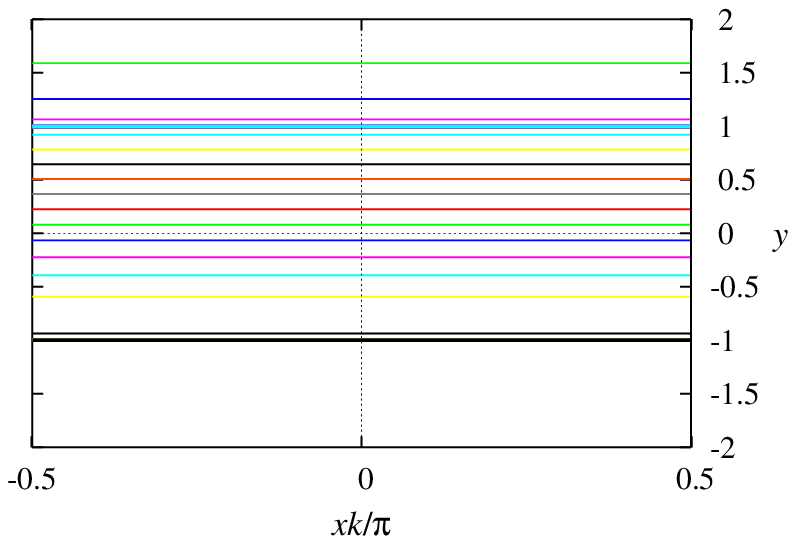}\put(-45,30){(\textit{e})}\includegraphics[bb=125bp 90bp 355bp 250bp,clip,width=0.49\columnwidth]{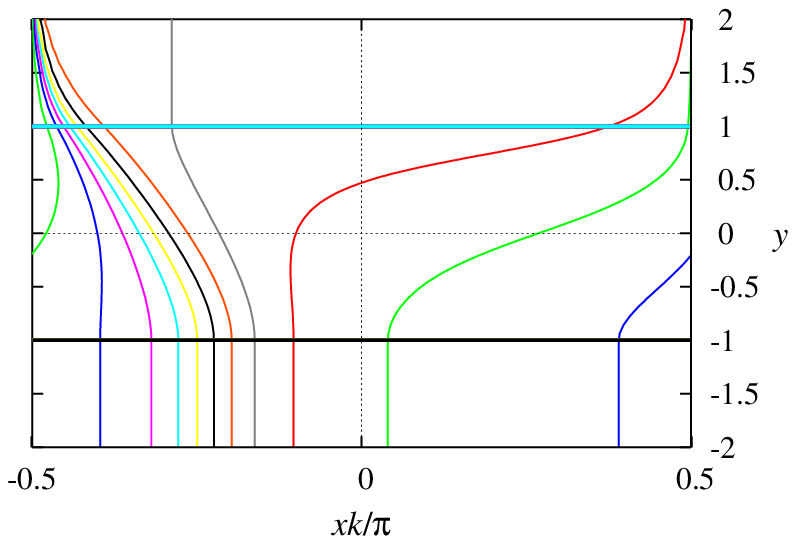}\put(-45,30){(\textit{f})}
\par\end{centering}

\caption{\label{cap:amp2-ph}The magnetic flux components in the phase with
the applied field (\textit{a,b}), shifted by $\pi/2$ (\textit{c,d}),
and the corresponding phase distribution (\textit{e,f}) for $\bar{\omega}=1$
and $k=1$ at $\Rm=0$ (layer at rest) (\textit{a,c,e}) and $\Rm=1$
(\textit{b,d,f}). }

\end{figure*}

The flux lines and the corresponding phase distribution of the vector
potential are plotted in figure \ref{cap:amp2-ph} for the layer at
rest (\textit{a,c,e}) and moving to the right with $\Rm=1$ (\textit{b,d,f}).
The flux components in the phase and shifted by $\pi/2$ relative
to the applied magnetic field correspond to the time instants when
the applied field is at maximum and absent, respectively. In the latter
case, the magnetic field is entirely due to the eddy currents. Figure
\ref{cap:amp2-ph}(e) shows that for the layer at rest, the phase
is constant over a half-wavelength of the applied field and varies
only with the vertical position except below the layer, where the
phase does not vary at all. It is important to note that this phase
distribution is actually piece-wise constant with the phase jumping
by $\pi$ across nodes of the standing wave. This phase discontinuity,
which is crucial for the subsequent analysis, is caused by two adjacent
halves of standing wave oscillating in opposite phases. As seen in
figure \ref{cap:amp2-ph}(f), this simple phase distribution breaks
down as soon as the layer starts to move. Although for a moving layer
the phase is no longer horizontally constant, it is still vertically
constant below the layer, where the field itself decays exponentially
with the vertical distance. This is an important result, which illustrates
why the phase measurements, in contrast to those of the amplitude,
are more robust and, in this case, actually independent of the vertical
position of the receiving coils. Note that such a perfect vertical
phase homogeneity holds only when the applied magnetic field is a
standing harmonic wave.

\begin{figure}
\begin{centering}
\includegraphics[width=0.5\columnwidth]{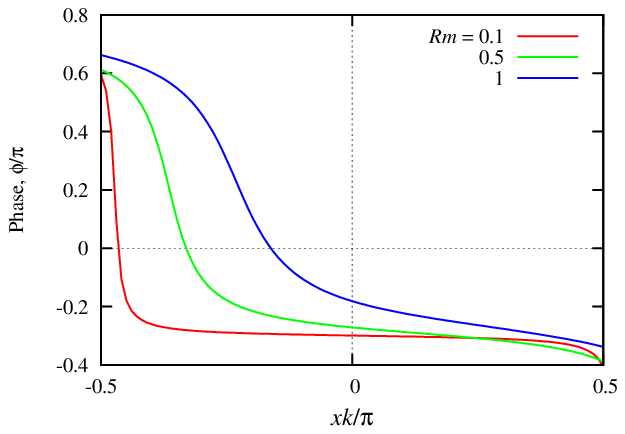}\put(-190,25){(\textit{a})}\includegraphics[width=0.5\columnwidth]{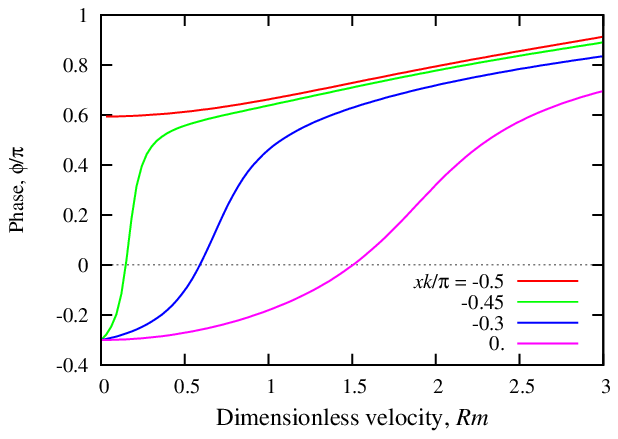}\put(-150,25){(\textit{b})}
\par\end{centering}

\caption{\label{cap:ph2-x}Phase distribution over a half-wavelength of the
applied magnetic field at various dimensionless velocities defined
by $\Rm$ (a) and the phase variation with $\Rm$ at different points
along the bottom of the layer for $\bar{\omega}=1$ and $k=1.$ }

\end{figure}

Figure \ref{cap:ph2-x}(\textit{a}) shows the phase distribution along
the bottom of the layer over a half-wavelength between two nodes of
the applied field for various dimensionless velocities $\Rm$. Henceforth
the phase $\varphi$ is presented in radians and scaled by $\pi$
so that $\pm\pi$ phase corresponds to $\varphi=\pm1.$ The original
phase discontinuity between adjacent half-waves shows up in \ref{cap:ph2-x}(\textit{a})
as soon as the layer starts to move. The increase in the velocity
is seen to smooth out this discontinuity and to shift it further downstream.
Note that the total phase variation over a half-wave remains $\pm1$
$(\pm\pi$) regardless of the velocity, as it should be for a spatially-periodic
solution.

The phase variation with the velocity at several observation points
along the bottom of the layer is plotted in figure \ref{cap:ph2-x}(\textit{b}).
As seen, the closer the observation point to the node, the steeper
the phase variation, but the shorter the velocity range of this variation.
The steep part of the phase variation with the velocity is obviously
due to the observation point lying in the transition region between
two adjacent half-waves discussed above. The phase variation is relatively
weak when the observation point is located either before or after
the transition region. This illustrates the importance of the location
of observation point, which for low velocities should be placed downstream
in close vicinity to the node or symmetry plane of the applied magnetic
field, where the phase varies significantly with the velocity. In
the case of the phase difference measured between two coils, the measurement
sensitivity can be increased by a horizontal offset of the sensing
coils with respect to the exciting coil so that one of the sensing
coils gets close to the midplane, as demonstrated experimentally in
the following.

\begin{figure*}
\begin{centering}
\includegraphics[width=0.5\columnwidth]{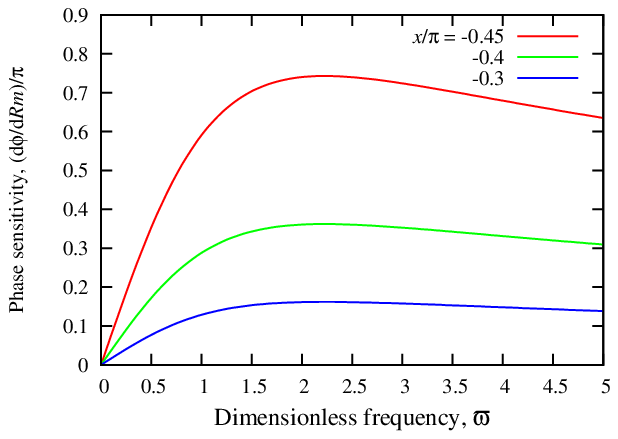}\put(-30,25){(\textit{a})}\includegraphics[width=0.5\columnwidth]{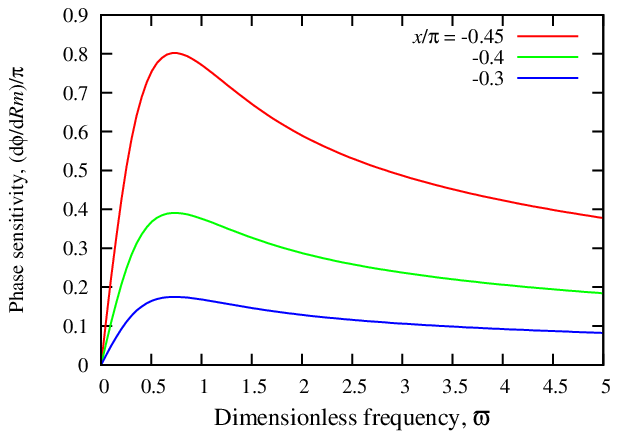}\put(-30,25){(\textit{b})}
\par\end{centering}

\caption{\label{cap:ph2-sns}The phase sensitivity versus the dimensionless
frequency $\bar{\omega}$ at various horizontal observation positions
below the layer for $k=1$ (\textit{a}) and $k=0.5$ (\textit{b}). }

\end{figure*}

To determine the optimal frequency of the applied ac field, it is
useful to consider the rate of variation of phase $\varphi$ with
the velocity $\Rm$. For sufficiently small $\Rm,$ which present
the main interest here, the phase sensitivity is defined as \begin{equation}
K=\frac{1}{\pi}\left.\frac{\partial\varphi}{\partial\Rm}\right|_{\Rm=0}.\label{eq:K}\end{equation}
 This quantity plotted in figure \ref{cap:ph2-sns} versus the dimensionless
frequency $\bar{\omega}$ shows that there is an optimal ac frequency
$\bar{\omega}_{o}$ at which the sensitivity attains a maximum. The
optimal frequency, which is seen to be independent of the horizontal
position of the observation point along the bottom of the layer, decreases
with the wavenumber of the applied magnetic field: $\bar{\omega}_{o}\approx2.1$
for $k=1$ and $\bar{\omega}_{o}\approx0.73$ for $k=0.5.$

\begin{figure}
\begin{centering}
\includegraphics[width=0.5\columnwidth]{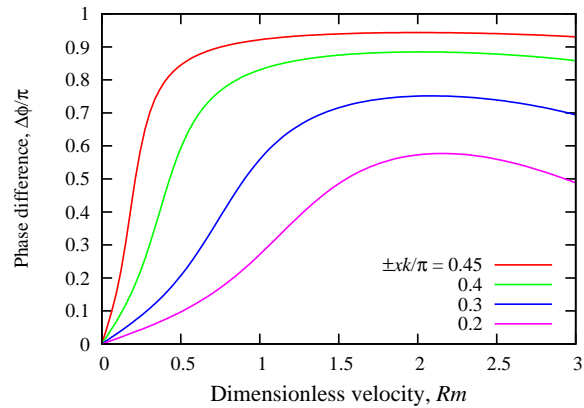} 
\par\end{centering}

\caption{\label{cap:ph2-dif}Phase difference between two observation points
placed symmetrically with respect to $x=0$ below the layer versus
the dimensionless velocity $\Rm$ for $\bar{\omega}=2$ and $k=1.$ }

\end{figure}

The phase difference between two observation points placed symmetrically
at various distances from $x=0$ below the layer is plotted in figure
\ref{cap:ph2-dif} versus $\Rm$ for $\bar{\omega}=2$ and $k=1.$
As discussed above, the velocity sensitivity of this phase difference
is seen to increase as the observation points are moved closer to
the nodes at $x=\pm0.5$. The closer the observation points to the
node, the higher the sensitivity but the shorter the velocity range
that can be measured. This is because the phase difference is seen
first to saturate and then to reduce due to smoothing by the motion
of medium. This smoothing effect limits the maximum velocity that
can be measured and becomes significant at $\Rm\gtrsim2,$ which corresponds
to rather high physical velocities.

\subsubsection{Sending coil modelled by two straight wires}

\begin{figure*}
\begin{centering}
\includegraphics[bb=125bp 90bp 355bp 250bp,clip,width=0.49\columnwidth]{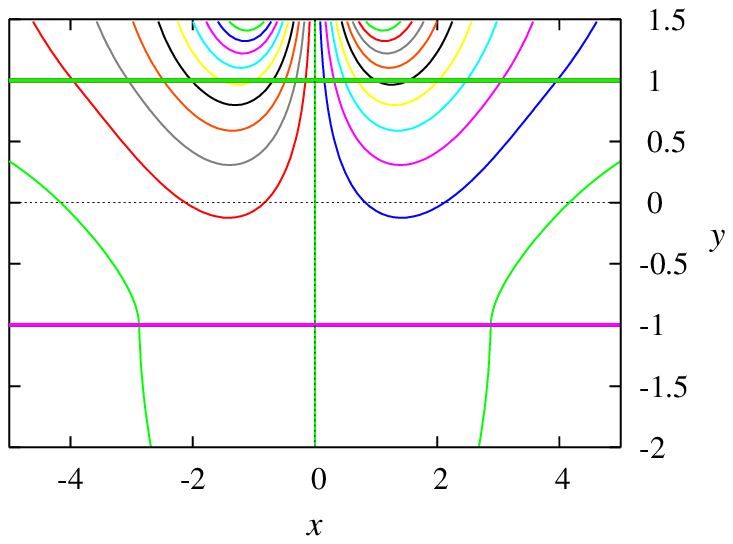}\put(-40,30){(\textit{a})}\includegraphics[bb=125bp 90bp 355bp 250bp,clip,width=0.49\columnwidth]{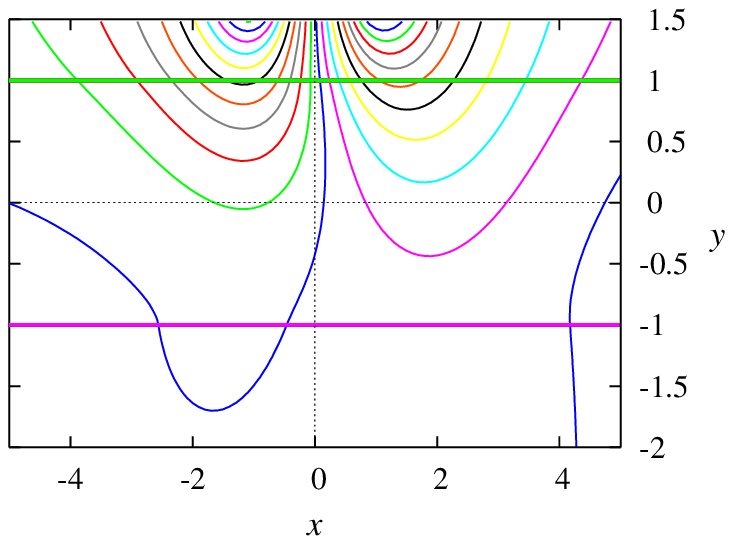}\put(-40,30){(\textit{b})}
\par\end{centering}

\begin{centering}
\includegraphics[bb=125bp 90bp 355bp 250bp,clip,width=0.49\columnwidth]{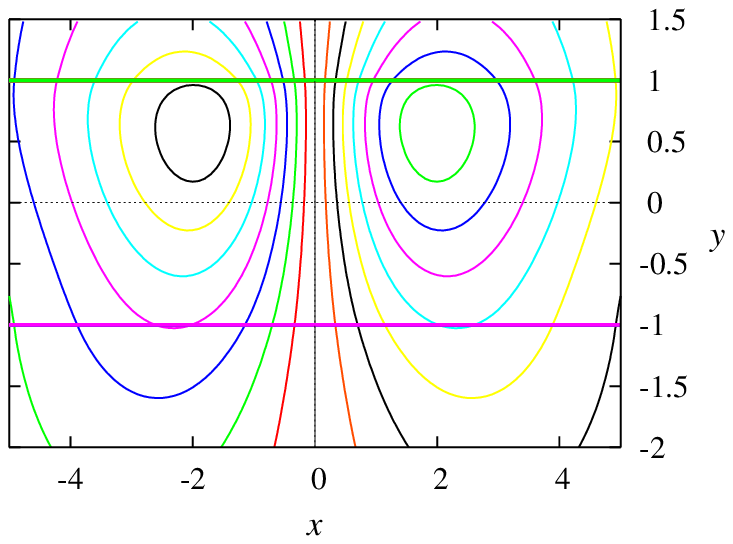}\put(-40,30){(\textit{c})}\includegraphics[bb=125bp 90bp 355bp 250bp,clip,width=0.49\columnwidth]{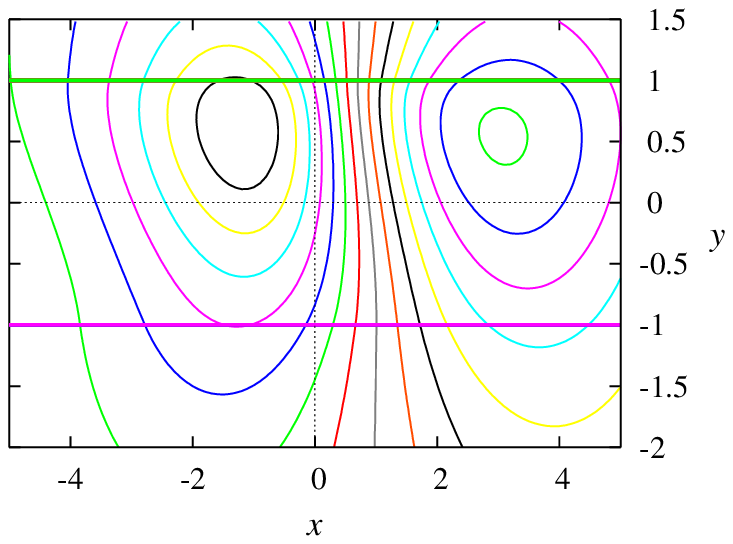}\put(-40,30){(\textit{d})}
\par\end{centering}

\begin{centering}
\includegraphics[bb=125bp 90bp 355bp 250bp,clip,width=0.49\columnwidth]{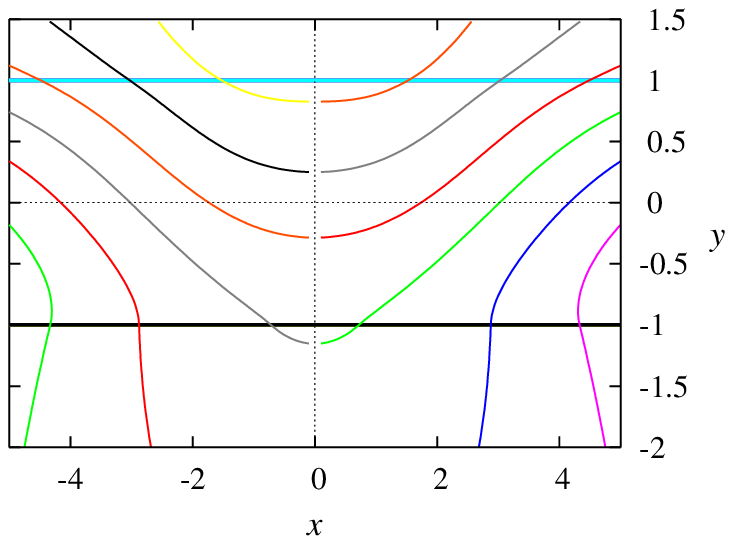}\put(-40,30){(\textit{e})}\includegraphics[bb=125bp 90bp 355bp 250bp,clip,width=0.49\columnwidth]{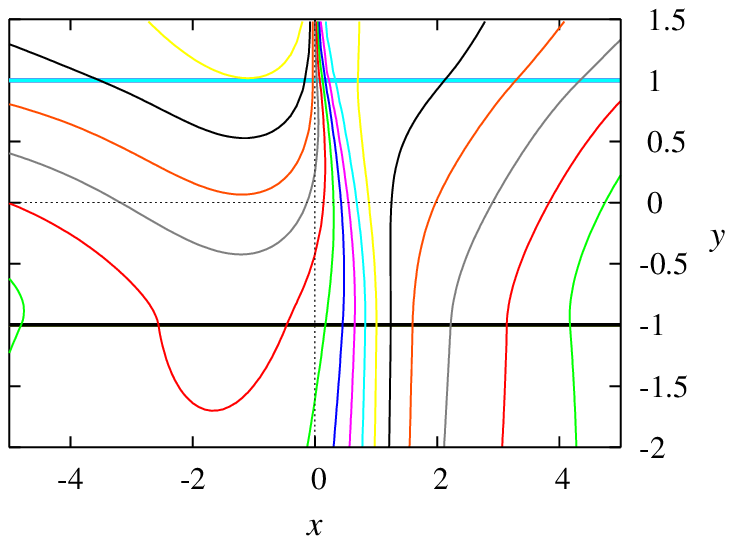}\put(-40,30){(\textit{f})}
\par\end{centering}

\caption{\label{cap:flwm2}The flux components in the phase with the applied
field (\textit{a,b}), shifted by $\pi/2$ (\textit{c,d}), and the
corresponding phase distribution (\textit{e,f}) for the external magnetic
field generated by two parallel wires with $\bar{\omega}=1$ at $\Rm=0$
(layer at rest) (\textit{a,c,e}) and $\Rm=1$ (\textit{b,d,f}). }

\end{figure*}

In this section, we turn to a more complicated external magnetic field
generated by a couple of parallel wires with opposite currents separated
by the horizontal distance $2s=2$ and put at the height $h=1$ above
the layer, as shown in figure \ref{cap:sketch}(b). The flux lines
and the corresponding phase distribution of the vector potential are
plotted in figure \ref{cap:flwm2} for the layer at rest (\textit{a,c,e})
and moving to the right with $\Rm=1$ (\textit{b,d,f}). In the former
case, the vector potential distribution is exactly anti-symmetric
with respect to the symmetry plane at $x=0,$ which thus is analogous
to a node in a standing wave. Correspondingly, there is a phase jump
of $\pi$ at $x=0$ when the layer is at rest. Figures \ref{cap:flwm2}(\textit{b,d,f})
show that the motion of the layer brakes the symmetry and smooths
out the phase discontinuity. Although, as seen in \ref{cap:flwm2}(\textit{f}),
the phase distribution below the layer is no more vertically invariant
as for the spatially-harmonic external magnetic field considered in
the previous section, its vertical variation is still weak in comparison
to that of the amplitude.

\begin{figure*}
\begin{centering}
\includegraphics[width=0.5\columnwidth]{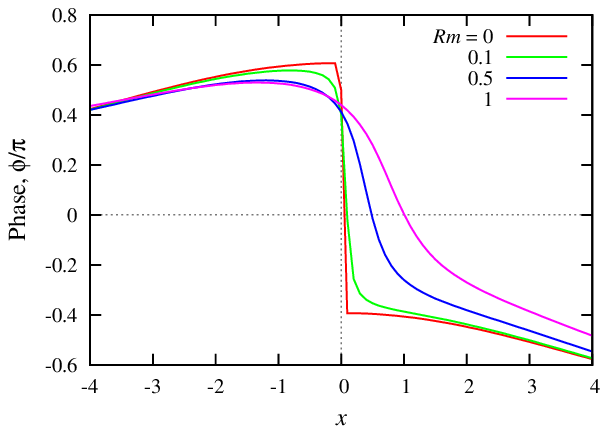}\put(-190,25){(\textit{a})}\includegraphics[width=0.5\columnwidth]{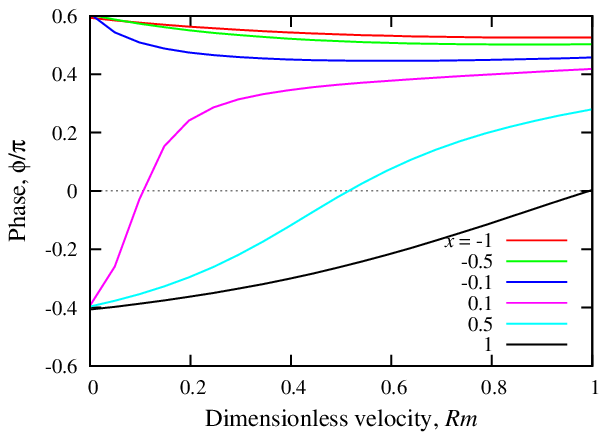}\put(-190,25){(\textit{b})}
\par\end{centering}

\caption{\label{cap:ph3-x}Phase distribution of the vector potential at various
velocities defined by $Rm$ (\emph{a}) and phase variation with the
velocity at different points along the bottom of the layer for $\bar{\omega}=1$
(\emph{b}). }

\end{figure*}

Figures \ref{cap:ph3-x}(a) and (b) show phase distributions at various
velocities $(\Rm)$, and the phase variation with $\Rm$ at different
points along the bottom of the layer for $\bar{\omega}=1$. In contrast
to the spatially-harmonic external magnetic field, now the phase distribution
is slightly non-uniform rather than piece-wise constant along the
layer even without the motion. Nevertheless, the field distribution
is still symmetric with respect to the midplane between the wires.
Motion is seen to break this symmetry and to smooth out the phase
jump. The phase variation with velocity is shown in figure \ref{cap:ph3-x}(b)
at several points along the bottom of the layer for $\bar{\omega}=1$.
It is important to note that sufficiently close to the symmetry plane,
the original phase non-uniformity at $\Rm=0$ is small relative to
that induced by the motion.

\begin{figure*}
\begin{centering}
\includegraphics[width=0.5\columnwidth]{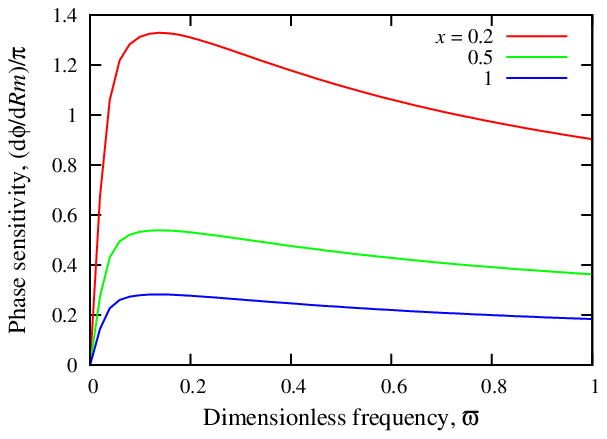}\put(-190,25){(\textit{a})}\includegraphics[width=0.5\columnwidth]{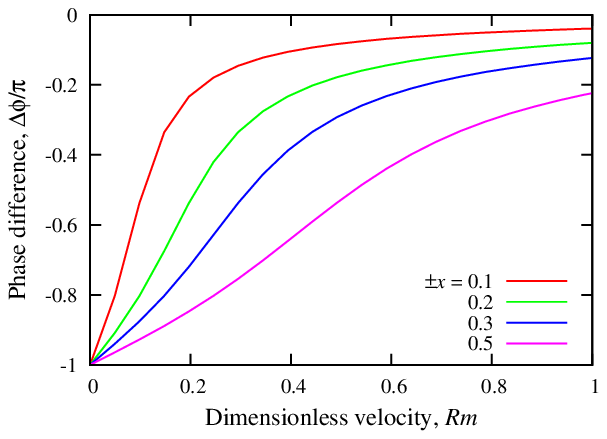}\put(-170,25){(\textit{b})}
\par\end{centering}

\caption{\label{cap:ph3-sns} The phase sensitivity versus the frequency at
various horizontal observation points at the bottom of the layer (\emph{a})
and the phase difference between two observation points placed symmetrically
with respect to $x=0$ at the bottom of the layer versus the dimensionless
velocity $\Rm$ for $\bar{\omega}=1$ (\emph{b}).}

\end{figure*}

The phase sensitivity introduced in the previous section, which is
plotted in figure \ref{cap:ph3-sns}(a) versus the dimensionless frequency,
shows that the optimal frequency at which the phase sensitivity attains
a maximum for this model is $\bar{\omega}\approx0.14.$ This frequency
is relatively low because in accordance to (\ref{eq:ck}) the applied
magnetic field is dominated by low-wavenumber (long-wave) modes. The
reduction of the sensitivity above the optimal frequency is rather
slow in comparison to its steep increase at sub-optimal frequencies.
Thus, the loss of the sensitivity at $\bar{\omega}\approx1$ is not
very significant, especially for the observation points further away
from the midplane. The phase difference between the pairs of observation
points placed symmetrically relative to the midplane, which is plotted
in figure \ref{cap:ph3-sns}(b) for $\bar{\omega}=1,$ shows the same
tendency as for the spatially harmonic external field. Namely, the
closer the observation points to the midpoint, the higher the sensitivity,
but the faster the saturation of the phase difference. Thus, the choice
of the observation points depends on the range of velocities to be
measured. In this case, when the observation points are placed exactly
symmetrically with respect to the midpoint, the original phase difference
at $\Rm=0$ is $\pi.$ This difference reduces and tends to zero with
the increase of $\Rm$ as the phase is smoothed out by the advection
of the magnetic field.

\section{\label{Realization}Flowmeter realization and test results}

\begin{figure*}
\begin{centering}
\includegraphics[width=0.29\textheight]{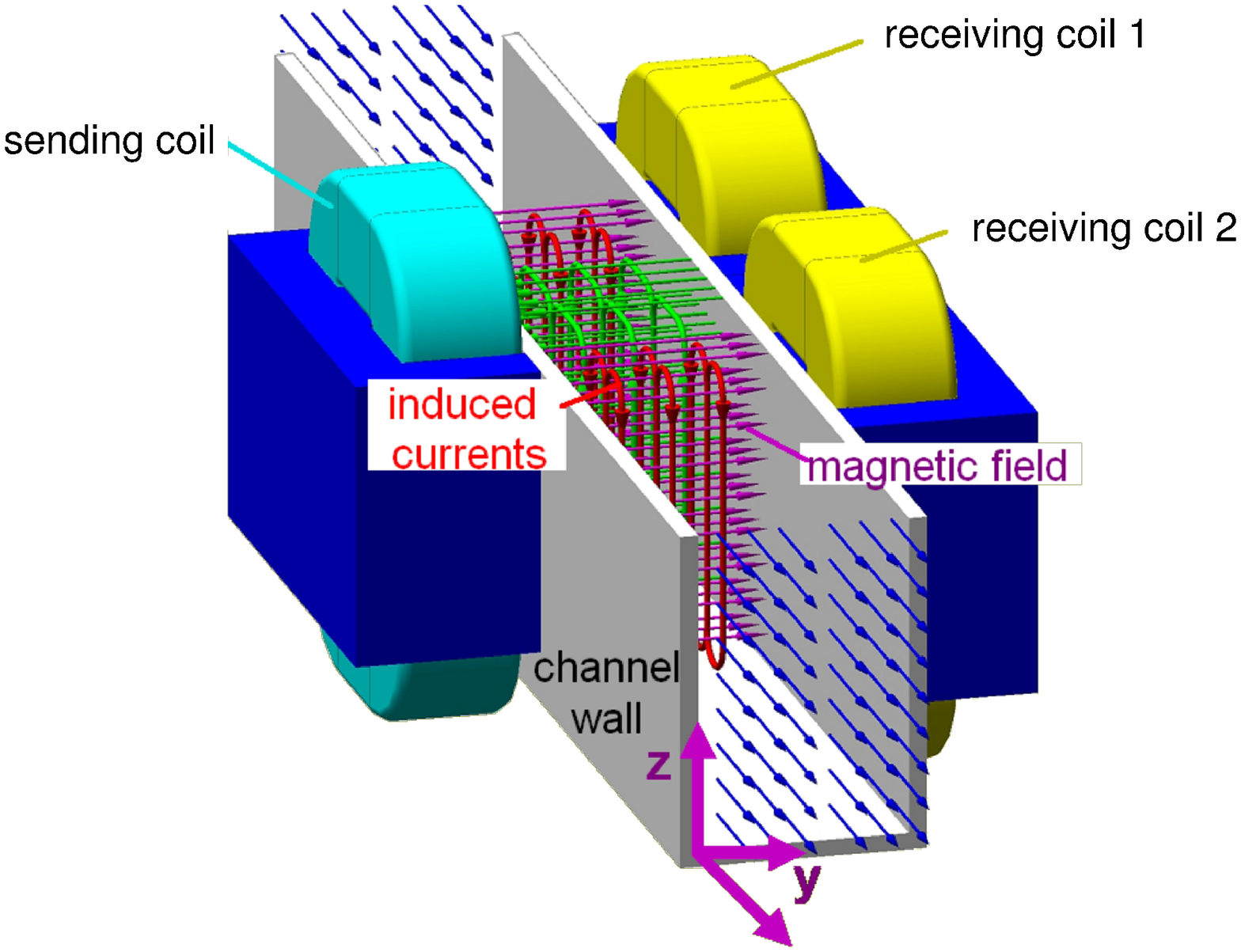}\put(-30,10){(\textit{a})}\includegraphics[width=0.29\textheight]{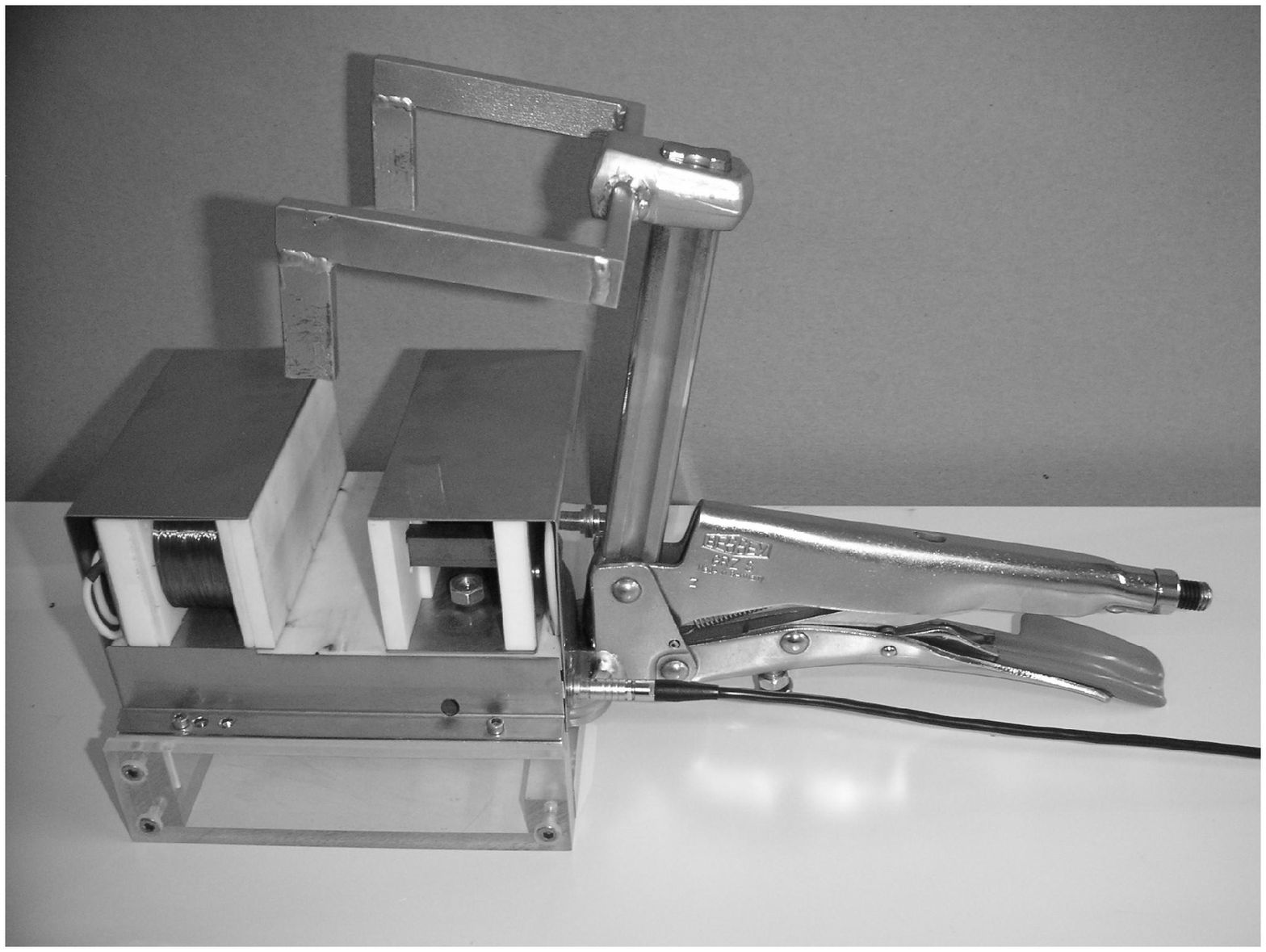}
\put(-30,10){(\textit{b})}
\par\end{centering}

\caption{\label{cap:Realisation}Experimental concept (\emph{a}) and a laboratory
model (\emph{b}) of the phase-shift flowmeter.}

\end{figure*}

A realization of the phase-shift flowmeter based on the principles
described above is shown in figure \ref{cap:Realisation}. It consists
of two receiving coils and a sending coil, which generates an ac magnetic
field, placed on the opposite sides of the duct with a liquid metal
flow. This flowmeter operates like a split transformer with two secondary
coils \cite{Sendai,Pat}. The phase shift between the voltages induced
in the two receiving coils is measured using a lock-in amplifier with
the internal averaging time of $100\,\mbox{ms}$ and the accuracy
of at least 2\%. 

\begin{figure}
\begin{centering}
\includegraphics[width=0.75\columnwidth]{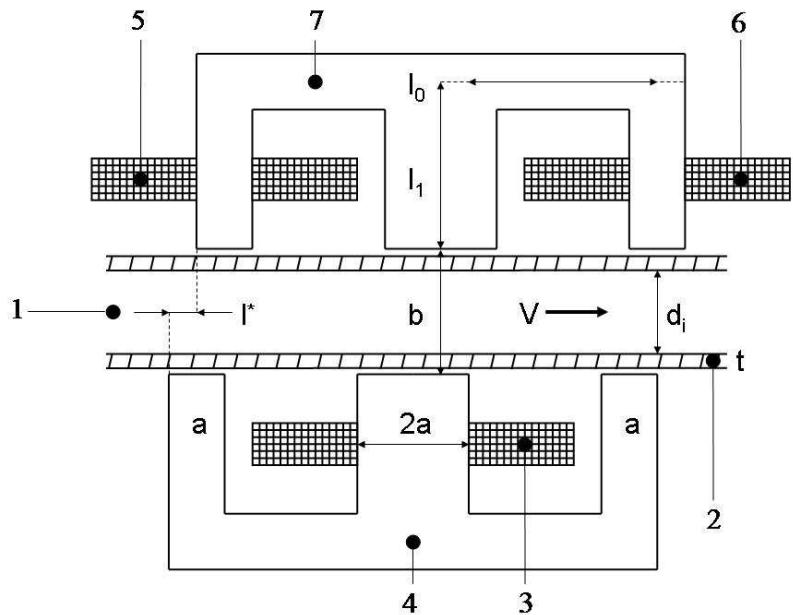} 
\par\end{centering}

\caption{\label{cap:scheme}The flowmeter setup with a liquid metal flow of
averaged velocity $V$ in pipe (2) of diameter $d_{i}$, sending coil
(3), laminated soft-iron yoke (4,7), receiving coils (5, 6), and the
horizontal offset $l^{\ast}$ between the receiving and sending coils,
which is zero for a symmetric but non-zero for asymmetric arrangements. }

\end{figure}

Receiving and sending coils can be placed either directly against
each other or shifted by some offset $l^{\ast},$ as shown in figure
\ref{cap:scheme}. Further we refer to these two arrangements as symmetric
and asymmetric ones. The theoretical analysis above suggests that
such an offset may enhance the sensitivity of the flowmeter.

\subsection{\label{Facilities}Test facilities}

The first facility is designed to operate with the eutectic melt of
GaInSn, which is liquid at room temperature and has the kinematic
viscosity $\nu=3.5\times10^{-7}\,\mbox{m/s}^{2}$ and electrical conductivity
$\sigma=3.3\times10^{6}\,\mbox{S/m}$ \cite{Mue-Bue}. Note that a
temperature increase by $1^{\circ}$ would cause the electrical conductivity
to drop by about $0.3\%,$ which according to (\ref{eq:Rm}) would
have the same effect as a corresponding reduction in the flow velocity.
The melt is driven by a permanent magnet induction pump \cite{Buc3}
with an adjustable flow rate. The lower part of the loop consists
of circular stainless steel tubes with inner diameter of $D=27\,\mbox{mm}$
and a wall thickness of $2.6\,\mbox{mm}.$ The upper part consists
of three independent test sections with the length of $400\,\mbox{mm}$
each---all with the same inner diameter and wall thickness. These
test sections can be opened and closed independently of each other
by valves. During experiments, two test sections were closed, which
ensured the cross-section-averaged flow velocity up to $V=1.4\, m/s$
in the third test section. This maximum velocity corresponds to $\Rm=0.08$
with the pipe radius taken for the length scale $H$. The corresponding
conventional Reynolds number $\RE=VD/\nu\approx10^{5}$ implies a
strongly turbulent flow. All test sections were kept completely filled
with the melt. The flow rate was independently controlled by a commercial
contact-type electromagnetic flowmeter (ABB, COPA-XL25) with 0.5\%
accuracy, whose operation was additionally verified by local Ultrasonic
Doppler Velocimetry measurements \cite{Cramer}. The commercial flowmeter
was also used to maintain a fixed flow rate by automatically controlling
the pump to compensate for the drop in the electrical conductivity
of the melt due to its ohmic heating by the pump.

The phase-shift flowmeter was attached to the tube of the GaInSn loop
by a stainless steel clamp seen in figure \ref{cap:Realisation}(b),
which allowed an easy installation on the tubes with a diameter up
to $34\,\mbox{mm}.$ The clamp arms were rounded in order to centre
the tube in the measurement gap of the flowmeter.

Additional experiments were carried out on a sodium loop \cite{Natan}.
Pipes and ducts were made of stainless steel ($\sigma_{w}=1.3\times10^{6}\,\mbox{S/m}$)
with the cross-sections of $45\times45$ and $45\times40\,\mbox{mm}^{2}$
(engineering tolerance $\pm0.2\,\mbox{mm}$) in the horizontal and
vertical test sections, respectively, which could be separated from
each other by valves. In order to avoid vibrations, in this case the
flowmeter was fixed on a separate frame rather than attached to the
pipe. At the operation temperature of $170^{\circ}$C, whose variation
was negligible during the measurements, the kinematic viscosity and
electrical conductivity of sodium were $\nu=5.5\times10^{-7}\,\mbox{m/s}^{2}$
and $\sigma=8.3\times10^{6}\,\mbox{S/m}$ \cite{Mue-Bue}. In this
case, an increase in temperature by $1^{\circ}$ would cause the electrical
conductivity and, thus, the apparent flow rate to drop by about $0.4\%.$
The flow was driven by an electromagnetic linear pump capable to produce
the flow rate up to $3\,\mbox{l/s},$ which was equivalent to the
cross-section-averaged velocity of $V=1.5\,\mbox{m/s}$ in the duct.
This maximum velocity corresponds to $\Rm=0.35$ with the duct half-width
$H=22.5\,\mbox{mm}$ taken as the length scale. The corresponding
Reynolds number $\RE=2VH/\nu\approx1.2\times10^{5}$ again implies
the flow to be a strongly turbulent.

Both facilities were equipped with calibrated Faraday-type electromagnetic
flowmeters. These reference flowmeters were used to calibrate and
test our phase-shift flowmeter, which needs to be re-calibrated when
applied to another flow configuration.

\subsection{Flow-rate measurements}

\label{Measurements}

\begin{figure}
\begin{centering}
\includegraphics[width=0.7\columnwidth]{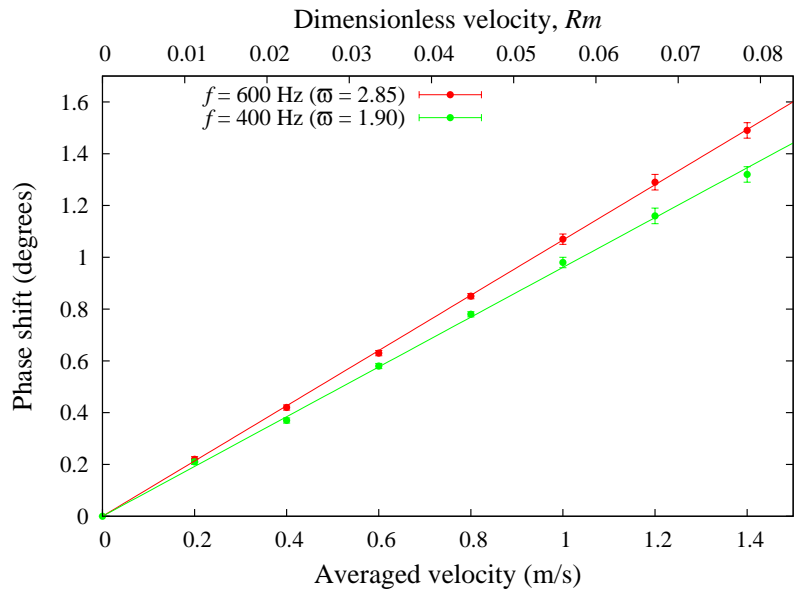} 
\par\end{centering}

\caption{\label{cap:InGaSn-v} The phase shift measured on the GaInSn loop
versus the flow-rate-averaged velocity in the symmetrically adjusted
case for $600\,\mbox{Hz}$ and $400\,\mbox{Hz}$ emitter frequencies,
which correspond to $\bar{\omega}=2.85$ and $\bar{\omega}=1.90$
dimensionless frequencies when the radius of the pipe $R=13.5\,\mbox{mm}$
is used as the length scale. The measured points are fitted with straight
lines crossing the origin.}

\end{figure}

\subsubsection{Flow-rate measurements on the GaInSn loop}

Figure \ref{cap:InGaSn-v} shows the phase shift measured on the GaInSn
loop depending on the cross-section-averaged velocity $V$ in a pipe
with insulating walls. For the emitter frequencies of $600\,\mbox{Hz}$
and $400\mbox{\,\mbox{Hz},}$ the phase shift is seen to vary nearly
linearly with the flow rate. The respective asymptotic standard errors
in the best linear fits are $0.3\%$ and $0.7\%.$ The corresponding
phase sensitivities $K,$ defined by (\ref{eq:K}), are $0.11$ and
$0.095.$ For the harmonic wave model, figure \ref{cap:ph2-sns} shows
that these sensitivities correspond to the observation point placed
at $x/\pi\approx-0.25,$ i.e. about an eighth of the wavelength downstream
from the node. For the emitter coil made of two straight wires according
to figure \ref{cap:ph3-sns}(a), the observation point has be located
at $x>1$, which is more than a half-distance between the wires downstream
from the midplane. Note that one cannot expect a quantitative agreement
with these very simple theoretical models, which are supposed to capture
only the basic effects but not particular details of the experiment.

\begin{figure}
\begin{centering}
\includegraphics[width=0.7\columnwidth]{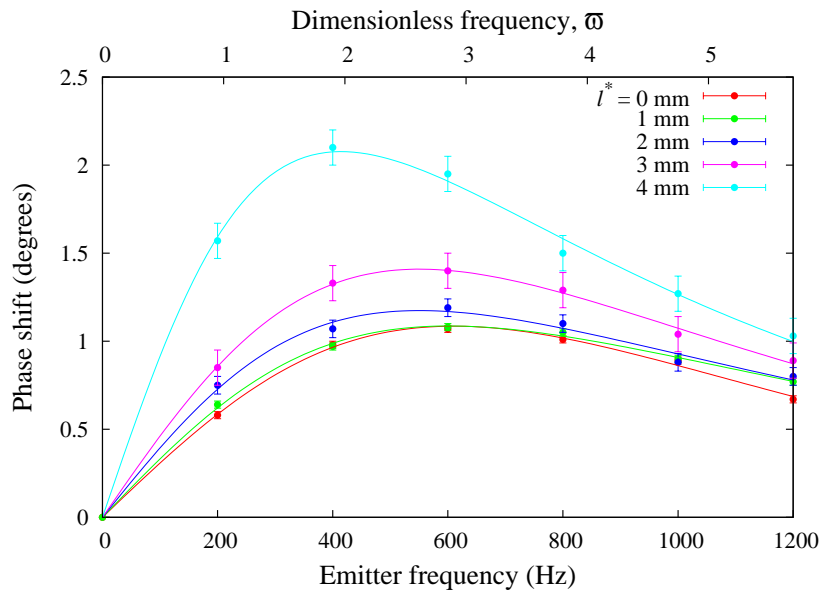} 
\par\end{centering}

\caption{\label{cap:InGaSn-f} Frequency response measured on the GaInSn loop
with insulating pipe walls for different receiving coil offsets at
the averaged flow velocity of $1\,\mbox{m/s}$.}

\end{figure}

To determine the frequency yielding the highest signal in the symmetric
and asymmetric flowmeter arrangements, the frequency response of the
flowmeter was investigated (see figure \ref{cap:InGaSn-f}). All experiments
were carried out under the same temperature, flow rate and velocity
profile of the melt. For the symmetric arrangement, the optimal frequency
was about $620\,\mbox{Hz}$, which corresponds to $\bar{\omega}=2.9$.
The optimal frequency decreased to about $400\,\mbox{Hz}$ when the
offset was increased to $l^{\ast}=4\,\mbox{mm}.$

\subsubsection{\label{Sodium}Measurements on the sodium loop}

The flowmeter used had a sending coil of 500 windings placed on the
one side of the duct and two receiving coils with 1000 windings each
placed on the other side of the duct. The sending coil was fed by
an alternating current in the range from a few tenths up to 3 A. Both
the sending and receiving coils were furnished with a high-permeability
laminated yoke. The coil wires were coated with two layers of high
temperature resistant polyamide ($T$ = 260$^{\circ}$C). Furthermore,
the coils were encased in the ceramic material MACOR, which can withstand
temperatures up to $800^{\circ}$C and, thus, protect the sending
and receiving coils from the hot pipe.

\begin{figure*}
\begin{centering}
\includegraphics[width=0.5\columnwidth]{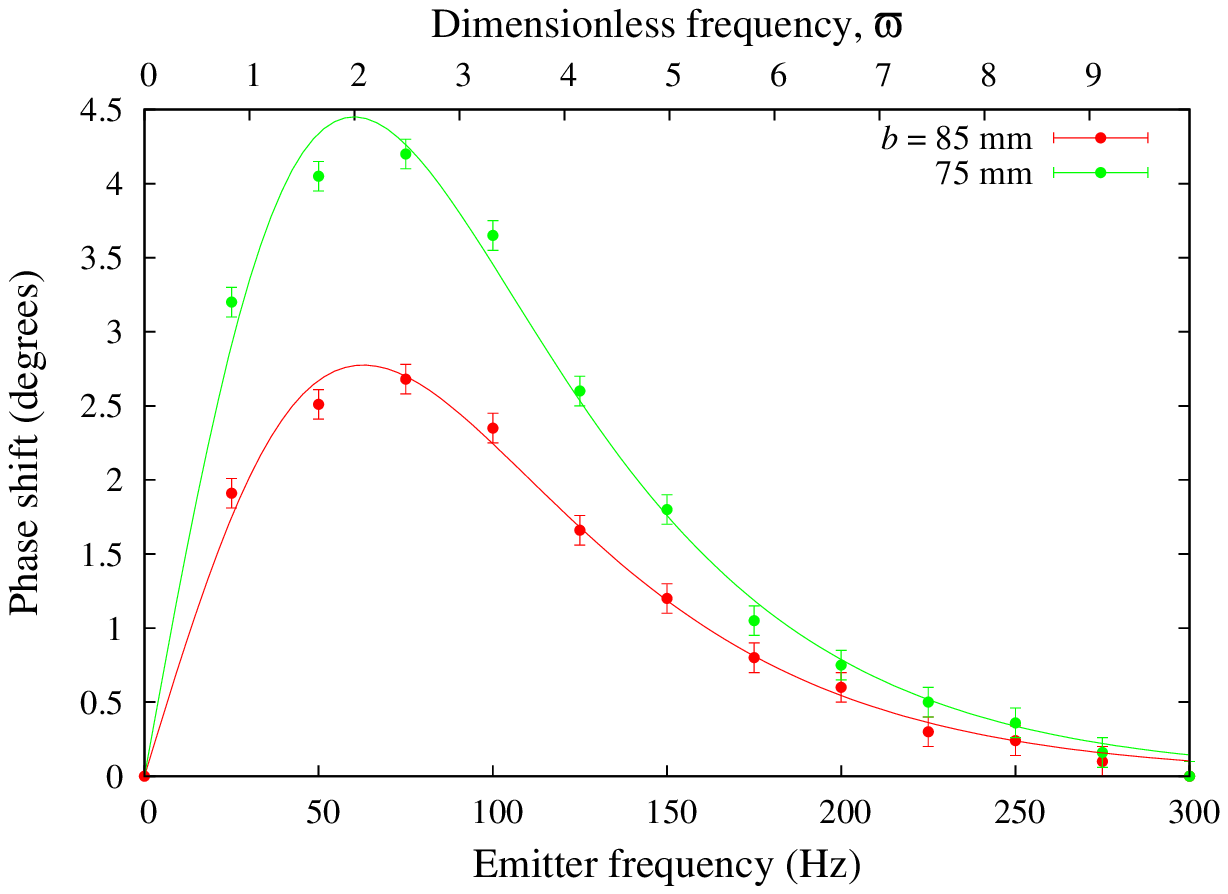}\put(-190,22){(\textit{a})}\includegraphics[width=0.5\columnwidth]{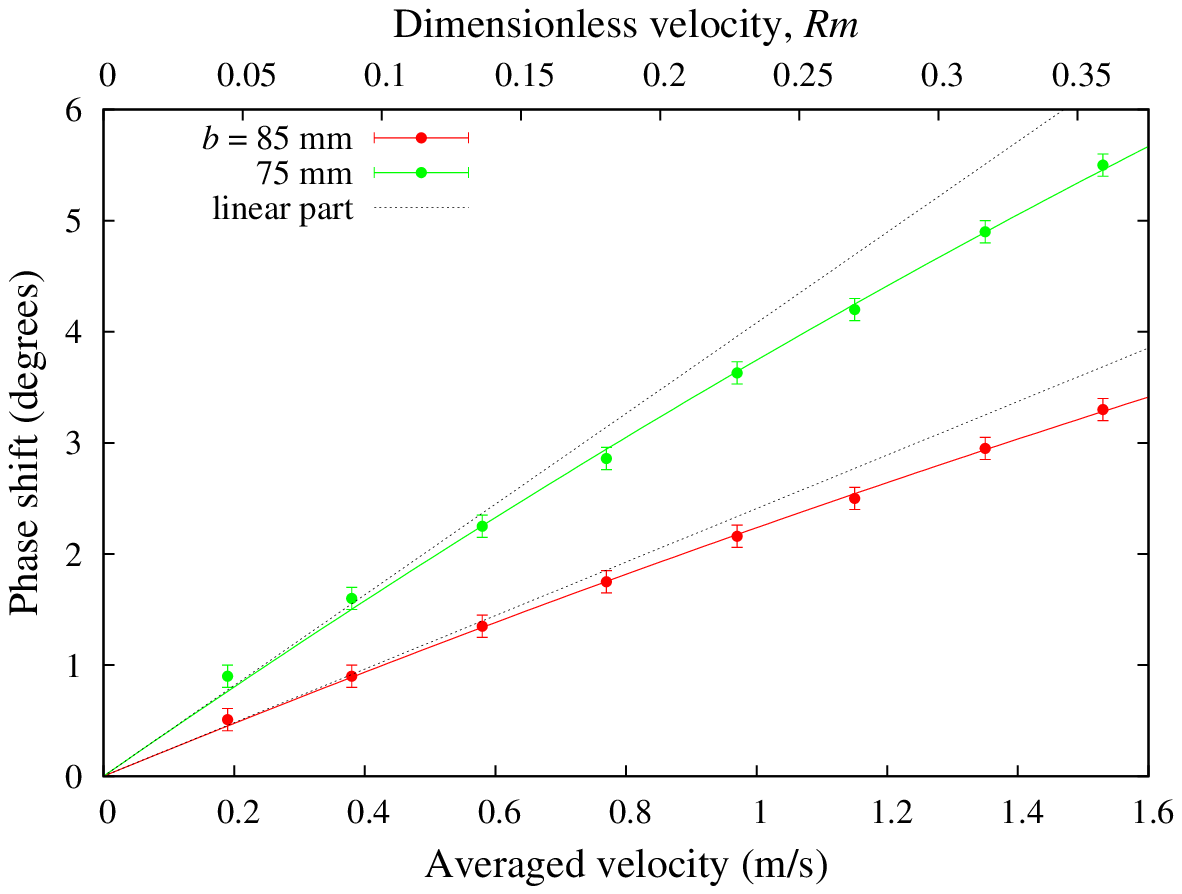}\put(-27,22){(\textit{b})} 
\par\end{centering}

\caption{\label{cap:Na-f-v}Measurements taken on the sodium loop with a symmetric
flowmeter arrangement for the coil gaps of $85\,\mbox{mm}$ and $75\,\mbox{mm}$:
(\emph{a}) the frequency response at the averaged velocity of $1.15\,\mbox{m/s}$
($\Rm=0.26)$ and (\emph{b}) the phase shift versus the  averaged
velocity at the emitter frequency of $75\,\mbox{Hz}.$ The latter
is fitted with a quadratic zero-crossing function and shown together
with its linear part.}

\end{figure*}

Figure \ref{cap:Na-f-v}(a) presents the phase shift measured at a
fixed flow rate depending on the emitter frequency for two different
withs of the gap between the sending and receiving the coils in a
symmetric arrangement (see figure \ref{cap:scheme}). These results
show the optimal frequency of about $70\,\mbox{Hz},$ which is much
lower compared to the GaInSn case due to the higher electrical conductivity
and the larger duct width. This optimal frequency corresponds to $\bar{\omega}=2.5$.
The phase-shift measurements presented in figure \ref{cap:Na-f-v}(b)
for two different widths of the coil gap show a noticeable deviation
from the linearity, which develops with the increase of the velocity
as the magnetic Reynolds number approaches $O(1).$ With the reduction
of the gap with from $85$ to $75\,\mbox{mm},$ the phase sensitivity
(\ref{eq:K}) raises from $K=0.057$ to $K=0.097,$ where the latter
is close to the value of $K$ corresponding to the measurements on
the InGaSn loop shown in figure \ref{cap:InGaSn-v}. 

\section{\label{Sum}Summary and conclusions}

In this paper, we have presented a conceptual design of a new contactless
ac induction flowmeter for liquid metal flows based on the phase shift
measurements. This flowmeter employs the fact that the flow of a conducting
liquid disturbs not only the amplitude but also the phase distribution
of an applied ac magnetic field. In order to figure out the basic
physical effects, we considered a simple model where the liquid flow
was approximated by a solid-body motion, and the conducting medium
was restricted to a layer of finite thickness. The sender coil was
approximated by by a spatially-harmonic standing magnetic wave in
one case and by a couple of straight wires placed above the layer
in another case.

In the case of an exciting magnetic field in the form of a standing
harmonic wave and no flow, the phase has a typical piece-wise constant
distribution along the layer with discontinuities at the wave nodes.
These discontinuities are smoothed out as soon as the flow sets in
and a strong phase gradient appears in the vicinity of the original
wave nodes. This phase variation can be measured to determine the
flow velocity. The closer the observation point to the node, the higher
the sensitivity of the phase to the velocity but the lower the velocity
at which the phase variation saturates. Thus, the higher the sensitivity,
the lower the maximal velocity, which can be measured. There is an
optimal ac frequency of the applied magnetic field which ensures maximal
sensitivity of the phase with respect to the velocity at small $\Rm.$
For the observation points below the layer, the optimal frequency
reduces with the increase of the wavelength of the applied magnetic
field. In the case of a spatially-harmonic standing magnetic wave,
the phase is independent of the vertical distance below the layer.

The phase distribution becomes more complicated when the magnetic
field is generated by a coil of finite size, which was modelled by
a couple of straight wires placed parallel to a conducting layer of
finite thickness. In this case, the phase is no longer uniform along
the layer even without the motion. However, there is still a characteristic
phase jump by $180^{\circ}$ at the midplane between the wires, while
the field itself is symmetric relative to this plane. The motion of
the layer smooths out this discontinuity and breaks the symmetry of
the phase distribution relative to the midplane. The asymmetry of
the phase distribution caused by the motion can be used to determine
the velocity. For this purpose, we consider the phase difference between
two observation points (receiver coils) placed symmetrically relative
to the midplane. The original phase non-uniformity at $\Rm=0$ becomes
relatively small when the observation points are placed sufficiently
close to the midplane, which is necessary for the detection of low
velocities. At higher velocities, the original phase non-uniformity
becomes negligible relative to that induced by the flow. Although
the optimal dimensionless frequency is rather low $(\bar{\omega}\approx0.14)$
for such a coil of finite size, where the exciting magnetic field
is dominated by long-wavelength modes, the decrease of sensitivity
with the frequency is slow. This allows one to work at frequencies
above the optimum at $\bar{\omega}\approx1$ without a significant
loss of sensitivity. The optimal distance of observation points (receiving
coils) from the symmetry plane depends on the velocity range to be
measured. A smaller distance ensures a higher sensitivity, which is
advantageous for lower velocities, but results in a reduced sensitivity
at higher velocities because of the saturation effect.

Based on these ideas a laboratory model of such a flowmeter was built
and tested on both GaInSn and sodium loops. A nearly linear relation
between the phase-shift and the flow rate was measured on the GaInSn
loop for the velocities up to $1.4\,\mbox{m/s},$ which corresponded
to small magnetic Reynolds numbers $\Rm\lesssim0.1.$ A noticeable
deviation from the linearity was observed to develop with the increase
of the velocity on the sodium loop up to $\Rm\approx0.3$5. The phase-shift
flowmeter was found to be a robust and accurate measurement device.
However, similar to standard eddy-current flowmeters, it is still
sensitive to the electrical conductivity and, thus, to the temperature
of the liquid.

\ack{}{This work was supported by Deutsche Forschungsgemeinschaft
in frame of the Sonderforschungsbereich 609, and by the European Commission
under contract FI6W-CT-2004-516520 in frame of the project EUROTRANS.}

\section*{References}

{}


\begin{thebibliography}{10}
\bibitem{JAS}Shercliff J A 1962 \textit{The Theory of Electromagnetic
Flow-Measurement} Cambridge University Press: Cambridge

\bibitem{Bevir}Bevir M K 1970 The theory of induced voltage electromagnetic
flowmeters \textit{J. Fluid Mech.} \textbf{43} 577--590

\bibitem{Engl_2}Engl W L, Arch. f\"ur Elektrotechnik 54 (1972) 269--277
(in German)

\bibitem{Shimizu} Shimizu T, Takeshima N and Jimbo N 2000 A numerical
study on Faraday-type electromagnetic flowmeter in liquid metal system
\textit{J. Nucl. Sc. Techn.} \textbf{37} 1038--1048

\bibitem{Duncombe}Duncombe E 1984 Some instrumental techniques for
hostile environments \emph{J. Phys. E: Instrum.} \textbf{17} 7--18

\bibitem{Buc1} Bucenieks I 2002 Electromagnetic induction flowmeter
on permanent magnets \textit{Proceedings 5th Int. PAMIR conference
on Fundamental and Applied MHD (Ramatuelle, France, September)} 103--105

\bibitem{Buc2} Bucenieks I 2005 Modelling of rotary inductive electromagnetic
flowmeter for liquid metal flow control. \textit{Proceedings 8th Int.
Symp. on Magnetic Suspension Technology (Dresden, Germany, September)}
204--208
 

\bibitem{Thess1}Thess A, Votyakov E V and Kolesnikov Y 2006 Lorentz
force velocimetry \textit{Phys. Rev. Lett.} \textbf{96} 164501

\bibitem{Thess2} Thess A, Votyakov E, Knaepen B and Zikanov O 2007
Theory of Lorentz force flowmeter \textit{New J. Physics} \textbf{9}
299

\bibitem{Priede1} Priede J, Buchenau D and Gerbeth G 2009 Force-free
and contactless sensor for electromagnetic flowrate measurements \textit{Magnetohydrodynamics}
\textbf{45} 451-458

\bibitem{PBG10}Priede J, Buchenau D and Gerbeth G 2010 Single-magnet
rotary flowmeter for liquid metals \emph{J. App. Phys.} (submitted)
arXiv:1012.3965 

\bibitem{LL48}Lehde H and Lang W T 1948 AC electromagnetic induction
flow meter \emph{US patent 2,435,043}

\bibitem{Cowley65}Cowley M D 1965 Flowmetering by a motion-induced
magnetic field \emph{J. Sci. Instrum.} \textbf{42} 406--409

\bibitem{Stefa} Stefani F, Gundrum Th and Gerbeth G 2004 Contactless
inductive flow tomography \textit{Phys. Rev. E} \textbf{70} 056306

\bibitem{Gund} Gundrum Th, Stefani F, Gerbeth G 2006 Experimental
aspects of contactless inductive flow tomography \textit{Magnetohydrodynamics}
\textbf{42}153--160

\bibitem{Feng} Feng C C, Deeds W E and Dodd C V 1975 Analysis of
eddy-current flowmeters \textit{J. Appl. Physics} \textbf{46} 2935-2940

\bibitem{Dement} Dementev S, Barbogallo F, Groesschel F, Bucenieks
I, Krysko S and Poznyaks A 2002 Preliminary LBE test of the electromagnetic
flow meter for the MEGAPIE target \textit{Magnetohydrodynamics} \textbf{38}
417--422

\bibitem{Sendai} Priede J, Buchenau D and Gerbeth G 2006 Contactless
electromagnetic induction flowmeter based on phase shift measurements
\textit{Proceedings EPM Conference (Sendai, Japan, October 23-27)}
735--740

\bibitem{Pat} Priede J, Gerbeth G, Buchenau D and Eckert S 2008 Verfahren
und Anordnung zur kontaklosen Messung des Durchflusses elektrisch
leitf\"ahiger Medien, \textit{German patent DE 102006018623B4}

\bibitem{Buc3} Bucenieks I 2000 Perspectives of using rotating permanent
magnets for electromagnetic induction pump design \textit{Magnetohydrodynamics}
\textbf{36} 181--187

\bibitem{Cramer} Cramer A, Zhang C and Eckert S 2004 Local flow structures
in liquid metals measured by ultrasonic Doppler velocimetry \textit{Flow
Meas. Instr.} \textbf{15} 145--153

\bibitem{Natan} Eckert S, Gerbeth G and Lielausis O 2000 The behaviour
of gas bubbles in a turbulent liquid metal magnetohydrodynamic flow
\textit{Int. J. Multiphase Flow} \textbf{26} 45--66 

\bibitem{Mue-Bue}Müller U, Bühler L 2001 \textit{Magnetofluiddynamics
in Channels and Containers}, Appendix A.1, Springer
\end{thebibliography}
\end{document}